\documentclass{article}

\usepackage{graphicx}
\usepackage{float} 
\usepackage{wrapfig}
\usepackage{amsmath}
\usepackage{amssymb}
\usepackage{braket} 
\usepackage[mathlines]{lineno}
\usepackage{ulem}
\usepackage{authblk}
\usepackage[margin=1in]{geometry}
\usepackage{url}
\usepackage[T1]{fontenc}
\usepackage[numbers,sort&compress]{natbib}

\begin{document}

\title{Artificial gauge fields in the t-z mapping for optical pulses: spatio-temporal wavepacket control and quantum Hall physics}

\author[1]{Christopher Oliver}
\author[2]{Sebabrata Mukherjee}
\author[3]{Mikael C. Rechtsman}
\author[4]{Iacopo Carusotto}
\author[1]{Hannah M. Price}

\affil[1]{School of Physics and Astronomy, University of Birmingham, Edgbaston, Birmingham, UK, B15 2TT}
\affil[2]{Department of Physics, Indian Institute of Science, Bangalore 560012, India}
\affil[3]{Department of Physics, The Pennsylvania State University, University Park, Pennsylvania 16802, USA}
\affil[4]{Pitaevskii BEC Center, INO-CNR and Dipartimento di Fisica, Università di Trento, I-38123 Trento, Italy}

\date{}

\maketitle

\begin{abstract}
We extend the $t-z$ mapping formalism of time-dependent paraxial optics by identifying configurations displaying a synthetic magnetic vector potential, leading to a non-trivial band topology in propagating geometries. We consider an inhomogeneous 1D array of coupled optical waveguides beyond the standard monochromatic approximation, and show that the wave equation describing paraxial propagation of optical pulses can be recast in the form of a Schr\"{o}dinger equation, including a synthetic magnetic field whose strength can be controlled via the transverse spatial gradient of the waveguide properties across the array. We use an experimentally-motivated model of a laser-written waveguide array to demonstrate that this synthetic magnetic field can be engineered in realistic setups and can produce interesting observable effects such as cyclotron motion, a controllable Hall drift of the wavepacket displacement in space or time, and unidirectional propagation in chiral edge states.
These results significantly extend the variety of physics that can be explored within propagating geometries and pave the way for exploiting this platform for higher-dimensional topological physics and strongly correlated fluids of light.
\end{abstract}

\section{Introduction}
A remarkable result of paraxial optics is that the electromagnetic field of paraxially-propagating classical light can be described by a Schr\"{o}dinger-like equation~\cite{Boyd2008}. In this equation, the usual time evolution of a wavefunction is replaced by the propagation of the electric field envelope along the optical axis, $z$, of the medium. In practice, a major platform for investigating paraxial propagation is arrays of coupled optical waveguides, laser-written into a substrate~\cite{Szameit2010}. In general, these propagating geometries can be used to emulate a variety of single-particle quantum phenomena~\cite{Schwartz2007, Levi2011, Verbin2013, Rechtsman2013a, Mukherjee2018, Szameit2011} or mean-field interacting physics if the medium is nonlinear~\cite{Segev1992, Eisenberg1998, Efremidis2002, Christodoulides2003, Fleischer2003, Lahini2008, Freedman2006}. This interacting physics includes Bose-Einstein condensates of photons, and opens the way to studies of fluids of light~\cite{Larre2015a,fontaine2020interferences,braidotti2022measurement}, with interesting perspectives towards quantum features~\cite{steinhauer2022analogue}.

One exciting avenue that has been explored intensively over the last 15 years is that of topological photonics~\cite{Ozawa2019, price2022roadmap}. In this field, the physics of topological phases of matter, originally discovered within the context of electrons in solids, is applied to photonics. Propagating geometries have proven to be a very fruitful platform in this context, with a major early achievement~\cite{Rechtsman2013b} being the investigation of Floquet topological insulators, in which a honeycomb array of waveguides acquires a non-trivial topology when a helical patterning of the waveguides is introduced. Since the propagation distance $z$ plays the role of the temporal evolution, the breaking of the translational symmetry along the $z$-direction of the helical waveguide system is analogous to the breaking of time-translation symmetry in a 2D tight-binding model of electrons under a temporally periodic modulation~\cite{lindner2011floquet}.
More generally, similar propagating geometries have proven to be a powerful tool for studying a wide variety of topological physics, including the investigation of the interplay between nonlinearity and topology~\cite{Lumer2013, Mukherjee2021, Mukherjee2020}, topological physics in non-2D geometries~\cite{Noh2017, Lustig2022, Yang2020, Fu2020b}, non-Hermitian effects in topology~\cite{Zeuner2015, Weidemann2020}, disorder~\cite{Meier2018, Stutzer2018}, Thouless pumping schemes~\cite{Kraus2012,zilberberg2018photonic}, transport~\cite{Wimmer2017} and quantum walks~\cite{Kitagawa2012}. 

So far, most if not all works on topological photonic effects using propagating geometries have employed monochromatic light, and so do not make significant use of the temporal dynamics associated with an optical pulse. From paraxial optics~\cite{Boyd2008}, it is well known that the Schr\"{o}dinger-like equation describing the propagation of an optical pulse in a dispersive medium also includes a second-order time derivative term, with the same structure as a kinetic energy term in quantum mechanics. This allows us to interpret time $t$ as an additional spatial dimension in addition to the transverse $x,y$ ones, while the coordinate $z$ along the propagation direction plays the role of a time. This interchange of the role of space and time is known as the $t-z$ mapping and has also been investigated at the quantum level in~\cite{Lai1989a,Lai1989b,Larre2015b,steinhauer2022analogue}.

In this work, we propose a novel configuration based on an array of coupled optical waveguides where a synthetic gauge field naturally appears when the temporal dynamics of an optical pulse is taken into account under the $t-z$ mapping. In particular, we consider propagation across a 1D array of coupled optical waveguides with slightly different properties, and show that the 2D wave equation for the classical optical field propagation in a mixed spatial-$j$/temporal-$t$ plane has the form of a Schr\"{o}dinger equation including a synthetic vector potential term as in the coupled wire model of quantum Hall physics~\cite{Kane2002, Teo2014, Budich2017}. A realistic configuration resulting in a sizable synthetic magnetic field and a non-trivial band topology is put forward, and experimentally accessible signatures of the magnetic effects are pointed out. These include a cyclotron motion of light wavepackets in the spatio-temporal $j-t$ plane, a Hall drift in response to additional synthetic electric fields in either the spatial or the temporal direction, as well as unidirectional propagation in chiral edge states. As compared to previous schemes~\cite{Ozawa2019,yuan2018synthetic} for synthetic magnetic fields and synthetic dimensions in arrays of microcavities~\cite{Ozawa2016,Yuan2016} or waveguides~\cite{dutt_creating_2022,wang_multidimensional_2020,hu2020realization,Piccioli2021}, our proposal has the crucial advantage that it does not require a dynamical modulation of the system and provides a new tool for the manipulation of the spatio-temporal shape of optical wavepackets. Moreover, the local interactions in our system suggest exciting prospects for strongly-correlated fluids of photons if the interaction strength can be scaled up.

The structure of the Article is the following: Sec.~\ref{sec:propagation_wpd} summarizes the mapping of the paraxial wave equation onto the Schr\"odinger equation with a synthetic magnetic field. The quantum Hall coupled wire model is reviewed in Sec.~\ref{sec:QHCW} and an experimentally realistic configuration for realizing it is presented in Sec.~\ref{sec:experimentally}. Observable signatures of the synthetic magnetic field and the non-trivial band topology are presented in Sec.~\ref{sec:observable}. Conclusions and perspectives towards quantum topological photonics and quantum fluids of light are finally sketched in Sec.~\ref{sec:conclu}.

\section{Mapping the paraxial wave equation onto a Schr\"{o}dinger equation with a synthetic magnetic field}
\label{sec:propagation_wpd}
In this first Section, we review the derivation of the well-known wave equation for the paraxial propagation of a pulse though an array of coupled waveguides~\cite{Boyd2008}. For suitably designed inhomogeneous arrays, we then map the wave equation onto a Schr\"{o}dinger equation for a particle in a synthetic magnetic field, where time plays the role of a synthetic spatial dimension and propagation through the array corresponds to time evolution. This equation will be our workhorse for the rest of the paper.

\begin{figure}[htbp]
	\includegraphics[width=\textwidth]{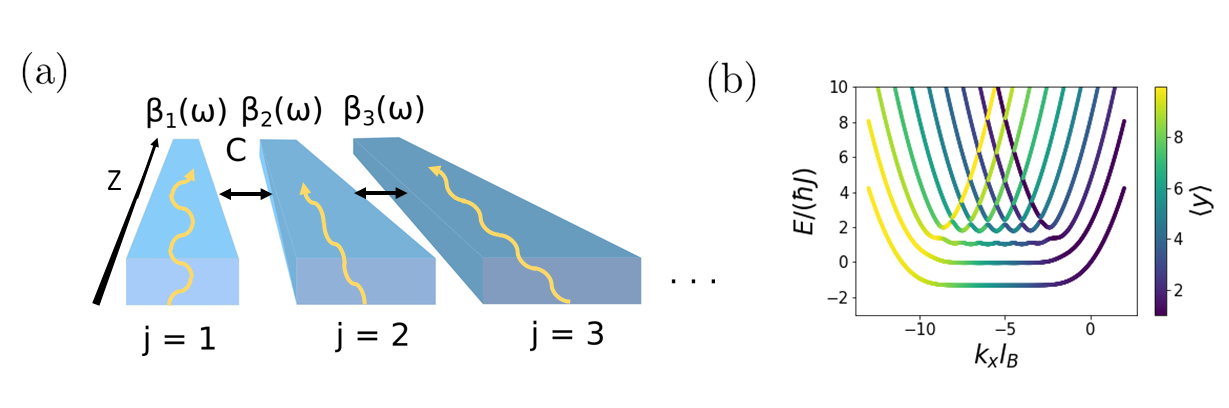}
	\centering
	\caption{(a): A schematic of our proposed setup consisting of a inhomogeneous 1D array of coupled single-mode waveguides. Each waveguide has a propagation constant $\beta_{j=1,\ldots,N}(\omega)$ and neighbouring waveguides are coupled together evanescently with a coupling strength $C$. The waveguide axis lies along the $z$-direction. The waveguide properties (e.g. the width and refractive index) vary across the array to engineer a non-trivial synthetic magnetic field. (b): Example of the band structure of a quantum Hall coupled wire model formed by $N = 10$ wires with periodic boundary conditions in the continuous $x$-direction. For each state of the band structure, the color coding indicates its average spatial position along $y$. We measure $k_x$ in units of $1/l_B$, where $l_B = \sqrt{\hbar / qB}$ is the magnetic length. The one free parameter of the Hamiltonian, the ratio of the two energy scales $r \equiv (\hbar^2/ml_B^2) / (\hbar J)$, is set to unity. We find, at low energies, dispersionless bulk Landau levels and chiral states localised on the system edge in the gaps, as is characteristic of a quantum Hall model.}
	\label{fig:F1}
\end{figure}

Consider an optical pulse propagating through a 1D array of $j=1,\ldots,N$ single-mode waveguides whose optical axis points in the $z$-direction, as shown schematically in Fig.~\ref{fig:F1}(a). We can write the electric field in waveguide $j$ as:
\begin{equation}
    E_j(\mathbf{r}, t) = {e}_j(x,y)a_j(z,t)e^{i(\beta_{j_{\text{ref}}}(\omega_0)z - \omega_0t)},
\end{equation}
where ${e}_j(x,y)$ is the electric field profile in the plane perpendicular to the optical axis; $\omega_0$ is the pulse carrier frequency and $\beta_{j_{\text{ref}}}(\omega_0)$ is the carrier propagation constant in the $j = j_{\rm ref}$ waveguide used as a reference. We choose this decomposition in order to to separate out the the envelope $a_j(z,t)$ which, within the paraxial approximation, is assumed to be slowly varying as a function of $z$ and $t$. We do not consider non-trivial polarisation effects so $E_j$ can represent any polarisation component of the electric field. We assume that the waveguides are effectively single-mode, meaning that any excited modes are well-separated from the fundamental mode in propagation constant, so that they play no role in the dynamics. We also neglect loss and disorder in the optical medium.

In frequency space, the propagation of the pulse is described by the coupled equations for the field amplitudes in the different waveguides:
\begin{equation}
    i\frac{\partial \tilde{a}_j}{\partial z} = -\left(\beta_j(\omega^{\prime} + \omega_0) - \beta_{j_{\text{ref}}}(\omega_0)\right)\tilde{a}_j - C\left(\tilde{a}_{j - 1} + \tilde{a}_{j + 1}\right),
    \label{eq:coupled_mode}
\end{equation}
where $\tilde{a}_j(z, \omega^{\prime})$ is the Fourier transform of $a_j(z,t)$ with respect to $t$ in terms of the frequency variable $\omega^{\prime} = \omega - \omega_0$ and $C$ is the evanescent coupling between neighbouring waveguides. For simplicity, this coupling is assumed to be frequency-independent in the range of interest and constant across the array. In typical implementations where the coupling strength is controlled by the spacing between waveguides, this latter condition may require an appropriate modulation of the spacing across the array to compensate for the variable size of the waveguides.

We now Taylor-expand $\beta_j(\omega)$ about $\omega_0$:
\begin{equation}
    \beta_j(\omega) \approx \beta_j(\omega_0) + \frac{d\beta_j(\omega)}{d\omega}\bigg\rvert_{\omega_0}(\omega - \omega_0) + \frac{1}{2}\frac{d^2\beta_j(\omega)}{d\omega^2}\bigg\rvert_{\omega_0}(\omega - \omega_0)^2,
    \label{eq:Taylor}
\end{equation}
where we neglect terms $O((\omega - \omega_0)^3)$ and higher. We then identify:
\begin{equation}
   {v_g^{(j)}} = \left[\frac{d\beta_j}{d\omega}\bigg\rvert_{\omega_0}\right]^{-1}, \qquad
    D_j = \frac{d^2\beta_j}{d\omega^2}\bigg\rvert_{\omega_0}, 
\end{equation}
as the group velocity and the group velocity dispersion in waveguide $j$, respectively. Substituting this expansion into Eq.~\ref{eq:coupled_mode} and Fourier-transforming back to the time domain produces the wave equation:
\begin{equation}
    i\frac{\partial a_j}{\partial z} = \frac{D_j}{2}\frac{\partial^2 a_j}{\partial t^2} - \frac{i}{v_g^{(j)}}\frac{\partial a_j}{\partial t} - \left(\beta_j(\omega_0) - \beta_{j_{\text{ref}}}(\omega_0)\right)\,a_j - C(a_{j + 1} + a_{j - 1}).
    \label{eq:wave_lab_frame}
\end{equation}
which can be easily mapped onto a Schr\"{o}dinger equation. 
To this purpose, we transform to a frame co-moving with the group velocity $v_g^{\rm (ref)} \equiv \left.d\beta_{j_\text{ref}}/d\omega\right|_{\omega_0}$ in the $j_{\rm ref}$ reference waveguide using a Galilean transformation with space and time interchanged, $\zeta \equiv z$ and $\tau \equiv t - z / v_g^{\text{(ref)}}$. After completing the square to eliminate the first time-derivative, Eq.~\ref{eq:wave_lab_frame} becomes:
\begin{equation}
    i\frac{\partial a_j^{\prime}}{\partial\zeta} = \frac{1}{2m_j}\left(-i\frac{\partial}{\partial\tau} - A_{j}^{(\tau)}\right)^2a_j^{\prime} + V_j\,a_j^{\prime} - C\left(a_{j + 1}^{\prime} + a_{j - 1}^{\prime}\right), 
    \label{eq:schrodinger}
\end{equation}
where we have defined
\begin{equation}
m_j = -\frac{1}{D_j}, \qquad
A_{j}^{(\tau)} = \frac{1}{D_j} \left(\frac{1}{v_g^{(j)}}- \frac{1}{v_g^{(\text{ref})}}\right),\qquad
V_j = \frac{1}{2D_j} \left(\frac{1}{v_g^{(j)}}-\frac{1}{v_g^{(\text{ref})}}\right)^2 - \left(\beta_j(\omega_0) - \beta_{j_{\text{ref}}}(\omega_0)\right)
\end{equation}
and where $a_j^{\prime}(\zeta, \tau)$ is the electric field envelope in the co-moving frame (denoted from now on by the prime symbol). 

This set of equations has the form of coupled Schr\"{o}dinger equations, where propagation along the optical axis of the waveguide array plays the role of time evolution, as we expect from the $t-z$ mapping. The particle with unit charge moves in a 2D system with one discrete ($j$) dimension  and one continuous ($\tau$) dimension: in the former direction, the hopping amplitude is $C$; in the latter, the particle has a (position-dependent) mass $m_j$ determined by the group velocity dispersion of the waveguides. 

On top of this, the particle experiences a magnetic vector potential in the $\tau$-direction $A_{j}^{(\tau)}$ and a scalar potential $V_j$. Crucially, the magnetic vector potential is oriented along $\tau$ and is waveguide-dependent, so non-trivial magnetic field effects can be engineered by introducing a spatial gradient of the waveguides' characteristics across the array. Since time-reversal is automatically broken by propagation, our proposal does not require any dynamical modulation of the system~\cite{Ozawa2021, Nemirovsky2021}. In contrast to models where a synthetic dimension is encoded in momentum-space quantities~\cite{Ozawa2021, Nemirovsky2021} or in the light frequency~\cite{Ozawa2016,Yuan2016,yuan2018synthetic,wang_multidimensional_2020,dutt_creating_2022,hu2020realization,Piccioli2021}, our proposed topological model is based on propagation in real space-time coordinates, which is of utmost interest in the long term to integrate local nonlinearities and realize strongly interacting photon models. 

We also note that, if one is not to make the Taylor expansion in Eq.~\ref{eq:Taylor} and wishes to keep the complete form of the waveguide dispersion $\beta_j(\omega)$, one obtains in the co-moving frame the following form of the evolution equation in Fourier space,
\begin{equation}
    i\frac{\partial \tilde{a}^{\prime}_j}{\partial \zeta} =  - \left(\beta_j(\omega^\prime + \omega_0) - \beta_{j_{\text{ref}}}(\omega_0)-\frac{\omega^\prime}{v_g^{\rm (ref)}}\right)\,\tilde{a}^\prime_j - C(\tilde{a}^\prime_{j + 1} + \tilde{a}^\prime_{j - 1})\,.
    \label{eq:wave_omega_lab_frame}
\end{equation}
For both this and our Schr\"{o}dinger equation (Eq.~\ref{eq:schrodinger}), the propagation eigenmodes of the array are then obtained by searching for stationary solutions of this equation in the form 
\begin{equation}
\tilde{a}^\prime_j(\zeta,\omega^\prime)=e^{i\,\Delta\beta^\prime\,\zeta}\,\tilde{a}^\prime_j(\omega^\prime)
\label{eq:coupled_mode_cm}
\end{equation}
where the propagation constant in the comoving frame, $\Delta\beta^\prime$, is related to the laboratory frame one by:
\begin{equation}
\beta(\omega)=\beta_{j_{\text{ref}}}(\omega_0) + \frac{\omega-\omega_0}{v_g^{\rm (ref)}}+\Delta\beta^\prime(\omega)\,.
\label{eq:mapping}
\end{equation}
For visualization purposes, in the following we will study the dispersion in terms of $\Delta\beta^\prime$, where $\Delta$ highlights that we consider propagation constants relative to a reference.

Finally, we note that, in the simplest limit where the mass is constant across the array ($m_j = m$), the scalar potential vanishes ($V_j = 0$), and the vector potential displays a linear gradient along $j$, $A_j^{(\tau)} = -\mathcal{B}j$, with $\mathcal{B}$ being a uniform magnetic field, this equation reduces to the well-known quantum Hall coupled wire model. In the next Section, we briefly move away from optics to review the physics of this model in general. This simple model will serve us as an intuitive guide for the following developments of the paper.

\section{The quantum Hall coupled wire model}
\label{sec:QHCW}
In the quantum Hall coupled wire model, a charged particle is subject to a uniform magnetic field and moves within a system of $N$ coupled wires: the particle can either freely move along each wire (as denoted by the continuous dimension, $x$) or hop between neighboring wires (along the discrete dimension, $y$)~\cite{Kane2002}. Hence, this model lies between the fully-continuous Landau levels for a particle in free space and the fully-discrete Harper-Hofstadter model for a particle on a 2D square lattice~\cite{Hofstadter1976}. Originally, the coupled wire model was investigated in the context of the fractional quantum Hall effect~\cite{Kane2002,Teo2014,Budich2017}; its interest is related to the ability to control the band flatness by varying the hopping between wires. Recently, it has also been realized experimentally using the internal atomic states as a (discrete) synthetic dimension in addition to a real spatial dimension~\cite{Chalopin2020}.

In mathematical terms, the coupled wire model is summarized by the Hamiltonian
{\begin{equation}
    \hat{\mathcal{H}} = \frac{\hbar^2}{2m}\left(\hat{k}_x + \frac{qB}{\hbar}y\right)^2 + \hbar J\sum_{y}(\ket{x, y + a}\bra{x, y} + \text{H.C.}),
    \label{eq:cw_ham}
\end{equation}}
where $q$ and $m$ are the particle's charge and mass along $x$ respectively, and $J$ is the hopping between adjacent sites along the discrete dimension $y$, of lattice spacing $a$. The magnetic field is uniform and equal to $B$, and a Landau gauge is adopted with the vector potential oriented along the continuous $x$ direction, ${\bf A}=-B y \hat{e}_x$. In our $t-z$-mapped Schr\"{o}dinger equation (Eq.~\ref{eq:schrodinger}), the waveguide index $j$ corresponds to $y$ and the time in the co-moving frame $\tau$ corresponds to $x$. The hopping $J$ corresponds to the evanescent coupling $C$ between neighbouring waveguides and the particle mass corresponds to the (reciprocal of the) group velocity dispersion. The magnetic vector potential $-By$ corresponds to our $A_{j}^{(\tau)}$. However, we emphasise that, as highlighted above, the quantum Hall coupled wire model is a general model that is of interest to several communities. We also note that, if we measure $k_x$ in units of the inverse of the magnetic length $l_B \equiv \sqrt{\hbar / qB}$ and the Hamiltonian in units of $\hbar J$, there is only one free parameter, namely the ratio of the two kinetic energy scales $r \equiv (\hbar^2/ml_B^2) / (\hbar J)$. 

To gain some intuition for the physics of the coupled wire model, we take periodic boundary conditions along the continuous ($x$) direction, as in~\cite{Kane2002}. This allows us to exploit the conserved momentum $k_x$ to diagonalise the Hamiltonian.
An example of the coupled wire model band structure calculated from the above procedure is shown in Fig.~\ref{fig:F1}(b), where the colouring of the states denotes their average position with respect to the discrete dimension. 

Physically, the most important features of this band structure are the existence, at low energies, of dispersionless bulk band states and of unidirectionally propagating edge states. The former are localised in the discrete bulk ({\it green/blue} colouring) and have an almost constant energy, corresponding to no group velocity; as such, they are a semi-discrete analog of the flat Landau levels of charged particles subject to homogeneous magnetic field in free space. The latter are localised on the edges of the system ({\it yellow/purple} colouring) and their $k_x$-dependent energy falls in the gaps between the flat levels; these states are one-way chiral edge states with non-zero group velocity and are protected by the non-trivial topology of the model, i.e. the non-zero Chern number of the bands. 

Intuitively, the appearance of these two types of states can be simply understood from Eq.~\ref{eq:cw_ham}. In the absence of the inter-wire coupling $J$, the dispersion consists of $N$ parabolae (corresponding to each of the $N$ wires) which are equally-spaced along $k_x$ due to the uniform magnetic field, and their minima have all the same energy. As energy increases, each parabola crosses sequentially with those of the neighbouring wires, except for the ones at the edges of the array, where neighbours are only present on one side. Turning on the inter-wire coupling $J$ then lifts the degeneracies around the crossings, mixing states and giving rise to the flat bulk bands in the center of the band structure and the localised one-way states at the edges that are visible in Fig.~\ref{fig:F1}(b). 

Having reviewed the physics of this general coupled wire model, we return to optics in the next Section, where we will see how the model can be naturally realized by the coupled Schr\"odinger equations (Eq.~\ref{eq:schrodinger}) in a suitably-designed waveguide array. We will also assess the impact of additional features such as the on-site potential $V_j$ and position-dependent mass $m_j$ terms. 

\section{An experimentally-motivated model of a laser-written waveguide array}
\label{sec:experimentally}
\begin{figure}
	\includegraphics[width=\textwidth]{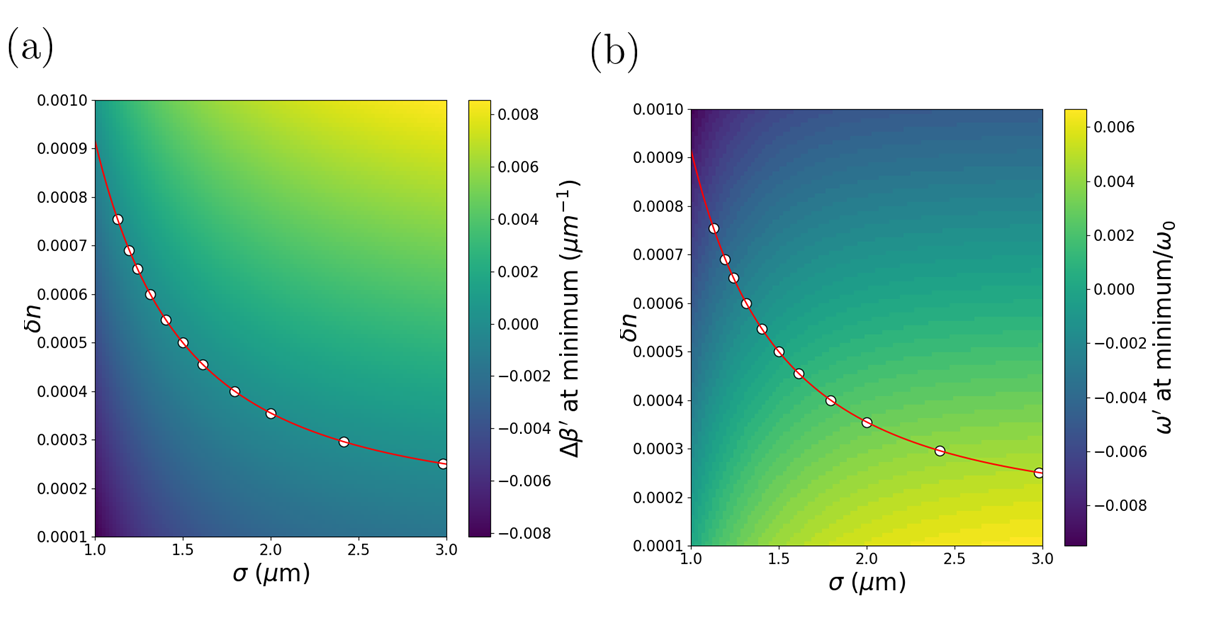}
	\centering
	\caption{Plots of (a): the co-moving frame propagation constant at the minimum of the waveguide dispersion and (b): the corresponding $\omega^{\prime}$ value, each as a function of the two main model parameters. We choose the minimum-$\Delta\beta^{\prime}$ = 0 contour ({\it red}) and choose points along it ({\it white}) that correspond to a uniform target spacing of $\Delta\omega^{\prime} = 0.001\omega_0$ $\omega^{\prime}$. The reference waveguide is located in the centre and has refractive index depth and width $\delta n_{\text{ref}} = 0.0005$ and $\sigma_{\text{ref}} = 1.5\mu\text{m}$ respectively. As it is discussed in detail in the text, material parameters are inspired from laser-written waveguides in fused silica glass and a carrier frequency corresponding to a wavelength of $0.5\mu$m is considered.}
	\label{fig:maps}
\end{figure}
From our discussion in the previous sections, the key ingredient to generate the synthetic magnetic field for photons is to design the $j$-dependence of the waveguide dispersion $\beta_j(\omega)$ in order to obtain a finite gradient along $j$ of the group velocity. To this purpose, we consider a model of $N$ waveguides embedded in a medium of frequency-dependent refractive index $n_0(\omega)$ and, for simplicity, we restrict our attention to a single transverse coordinate $x$. Within each waveguide $j$, light is confined by a (frequency-independent, for simplicity) lateral spatial profile of the refractive index. More precisely,
{\begin{equation}
    n_j(x, \omega) = n_0(\omega) + \delta n_j \exp\left(-\left(\frac{x^2}{2\sigma_j^2}\right)^m\right), 
    \label{eq:ref_index}
\end{equation}}
where for concreteness and with no loss of generality we consider the specific example of the refractive index $n_0(\omega)$ of fused silica glass~\cite{datasheet} used in many recent experiments~\cite{ozawa2019topological}. Experimentally-motivated $m = 10$ and $\delta n_j > 0$ values are taken for the super-Gaussian power of the spatial profile and the refractive index shift respectively. 

In order to obtain the synthetic magnetic field, different values of the width $\sigma_j$ and the refractive index depth $\delta n_j \ll 1$ must be taken for the different waveguides. In experiments, these parameters are controlled by varying the speed at which the optical medium is translated across the beam used for writing.
\begin{figure}
	\includegraphics[width=\textwidth]{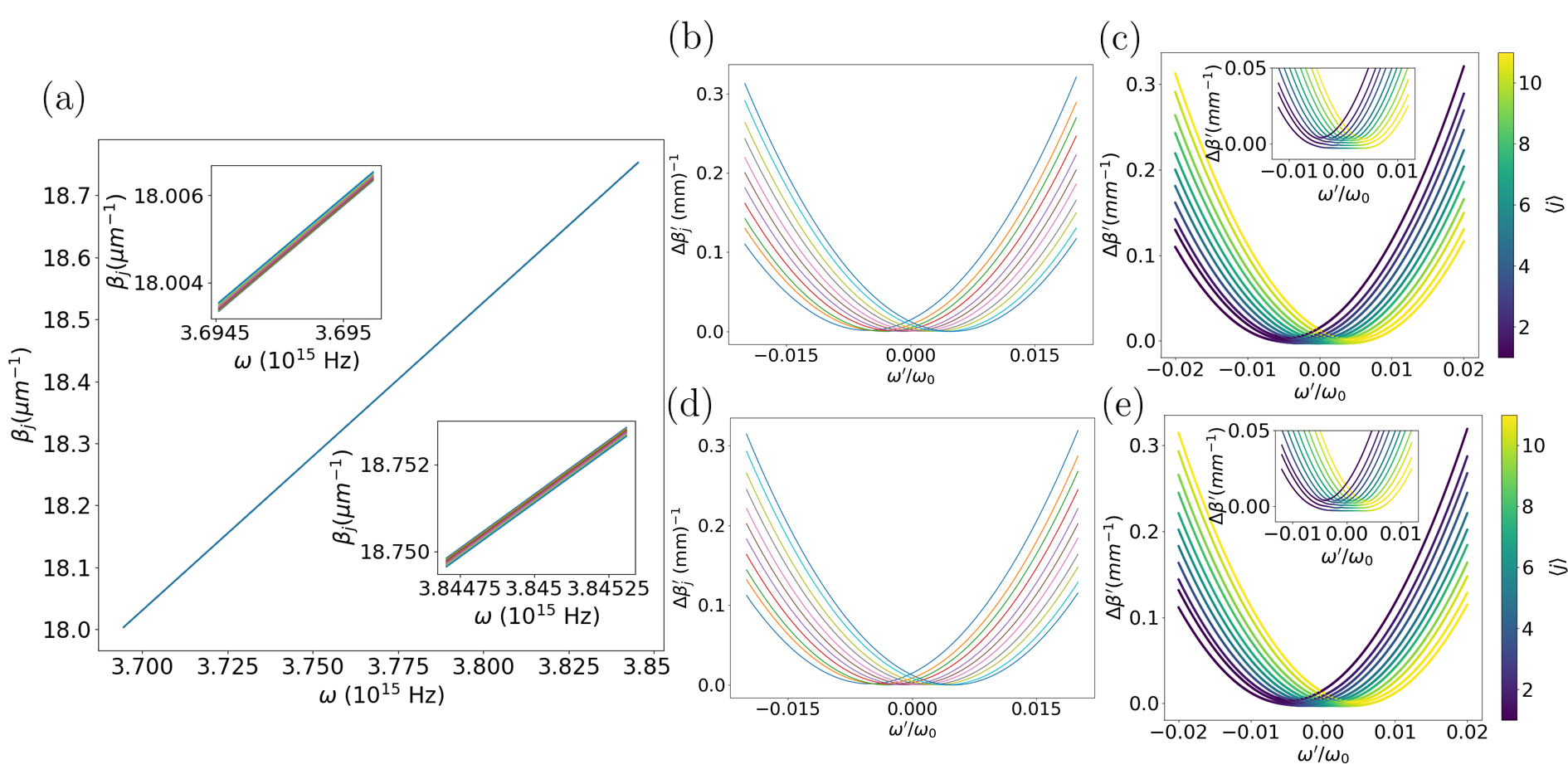}
	\centering
	\caption{(a): Propagation constants in the lab frame for an array of $N = 11$ uncoupled waveguides with refractive index profile as in Eq.~\ref{eq:ref_index}, after the waveguide widths and refractive index perturbations have been tuned to make the co-moving frame coupled array band structure have similar features as the quantum Hall coupled wire model. The insets show two different frequency ranges, with the order of the curves reversed between the two, showing that the curves intersect each other. (b): The uncoupled propagation constants in (a) transformed into the co-moving frame as described in the text. We find minima at approximately the same propagation constant value and approximately uniform spacing in frequency. Physically, these correspond to having an almost-constant scalar potential and a near-uniform magnetic field respectively. (c): The dispersion curves in (b) when a nearest-neighbour evanescent coupling of $C = -0.002$mm$^{-1}$ is included, showing avoided crossings. The colour of each state denotes the expectation value of its position with respect to the discrete direction. We see that this band structure includes bulk states ({\it green}) that are nearly dispersionless (see inset) and chiral edge modes ({\it purple/yellow}) within the gap like in the coupled wire model. (d) and (e): The uncoupled [(d)] and coupled [(e)] co-moving frame dispersions calculated from our Schr\"{o}dinger equation (Eq.~\ref{eq:schrodinger}), showing excellent agreement with (b) and (c). Throughout this figure, the parameters are as in Fig.~\ref{fig:maps}.}
	\label{fig:F2}
\end{figure}
We calculate the dispersions $\beta_j(\omega)$ for our refractive index profile by numerically solving the Helmholtz equation for our refractive index profile~\cite{Boyd2008,Ozawa2019}:
\begin{equation}
    i\frac{\partial \tilde{a}_j(x, z, \omega)}{\partial z} = -\frac{c}{2n_0(\omega)\omega}\frac{\partial^2 \tilde{a}_j(x, z, \omega)}{\partial x^2} - \frac{\omega}{c}\delta n_j \exp\left(-\left(\frac{x^2}{2\sigma_j^2}\right)^m\right)\tilde{a}_j(x, z, \omega),
\end{equation}
which has the form of a Schr\"{o}dinger equation in which the refractive index perturbation plays the role of a potential well. We write $\tilde{a}_j(x,z,\omega) = \tilde{a}_j(x, \omega)e^{i\delta\beta_j(\omega)z}$, where $\delta\beta_j(\omega)$ is the part of the propagation constant due to the waveguide itself, and we diagonalise the resulting equation. We choose the fundamental mode, and verify that the other modes are well-separated in propagation constant. This produces the total dispersion $\beta_j(\omega) \equiv n_0(\omega)\omega / c + \delta\beta_j(\omega)$. We can then employ our mapping (Eq.~\ref{eq:mapping}) to change to the co-moving frame.

We then need to adjust our free parameters of the array, $\delta n_j$ and $\sigma_j$, to make our co-moving frame dispersion curves as close as possible to the coupled wire model, i.e. a uniform horizontal spacing between the curves corresponding to a uniform magnetic field, and the minima of the curves all being level vertically, corresponding to no on-site potential. To do this, we sweep out the $(\sigma, \delta n)$ parameter space and, for each point in the space, we find the dispersion $\beta(\omega)$  in the above way. 
We choose the co-moving frame to be defined by the reference waveguide with parameters $\delta n_{\text{ref}} = 0.0005$ and $\sigma_{\text{ref}} = 1.5\mu\text{m}$, and we choose a carrier frequency $\omega_0$ corresponding to a wavelength of $0.5\mu$m. Within this comoving frame, we identify for each value of the model parameters $(\sigma, \delta n)$ the minimum of the dispersion. We then plot the values of the $\Delta\beta^{\prime}$ propagation constant and of the $\omega^{\prime}$ frequency at this minimum as a function of $(\sigma, \delta n)$. The results of this are shown in Fig.~\ref{fig:maps}. 

To select the parameters for our array, we choose the zero contour of the minimum propagation constant surface in order to force all the curves to have their minima at the same value, corresponding to a vanishing on-site potential. We then sample $N$ points from the chosen contour for an array of $N$ waveguides, placing the reference waveguide in the centre (the contour is shown in red, with the sampled points in white). We choose the points to enforce a chosen frequency spacing, $\Delta\omega^{\prime}$, between adjacent dispersions (i.e. a chosen magnetic field strength). The end results in the lab frame are plotted in Fig.~\ref{fig:F2}(a), and in the co-moving frame in Fig.~\ref{fig:F2}(b).

Including the evanescent coupling $C$ into our un-coupled dispersion and diagonalisation of the co-moving frame equation (Eq.~\ref{eq:wave_omega_lab_frame}) gives the eigenmodes shown in panel Fig.~\ref{fig:F2}(c) and, in a magnified view, in the inset of this panel. The qualitative resemblance with the quantum Hall coupled wire model is apparent: the bottom of the dispersion forms isolated bands corresponding to almost flat Landau levels that transform into edge states on the sides of the dispersion. The color scale highlights the spatial location of the different states: as expected, Landau levels are localized in the bulk, while the chiral edge states sit on the extreme waveguides $j = 1$ ({\it purple}) and $j = N$ ({\it yellow}). 

For comparison, we also calculate the band structure for the waveguide array using our Schr\"{o}dinger equation (Eq.~\ref{eq:schrodinger}). To this purpose, we use our lab frame waveguide dispersions $\beta_j(\omega)$ to calculate the effective mass, scalar on-site potential and magnetic vector potential around the reference waveguide and carrier frequency, as shown in the Supplemental Material~\cite{supmat}. We then diagonalize the Schr\"odinger equation (Eq.~\ref{eq:schrodinger}) for different values of $\omega^\prime$ to find the propagation constants in the comoving frame. The results for no coupling ($C = 0$) and for a finite coupling $C$ are shown in Fig.~\ref{fig:F2}(d) and (e) for the same parameters as for panels (b) and (c). The agreement between the two calculations is excellent, which further confirms the power of our configuration to generate a non-trivial synthetic magnetic field and thus realize a topological quantum Hall coupled wire model. As an aside, we mention that the engineering of the photonic band structure to have quantum Hall features is not unique to this model. In the Supplemental Material, we present an analytically-solvable toy model, whose band structure we also tune to resemble the coupled wire model~\cite{supmat}. In the next section we will investigate observable signatures of the synthetic magnetic field on optical quantities of experimental interest.

We note that the couplings used throughout this work are $C \sim 0.001$mm$^{-1}$ in magnitude and require correspondingly long waveguides or state recycling techniques~\cite{Mukherjee2018b} to observe any dynamics, as discussed in the next section. While such a regime could be obtained in experiment by the use of sufficiently large inter-waveguide spacings and long glass samples, the system might turn out to be sensitive to disorder in the optical medium, which we do not include in our model. If required, the analytical treatment in Sec.~\ref{sec:propagation_wpd} suggests several strategies to overcome the difficulty: the viable range of $C$ and the required waveguide length are in fact determined by the characteristic ``kinetic energy'' $\Delta \beta_{\rm char}'\sim D_j \Delta\omega'^2$ in the temporal direction, which is determined by the bandwidth of the light source (here taken to be $\Delta\omega'\sim 10^{-3}\omega_0$) and by the group velocity dispersion $D_j$ of the waveguides. The former can be increased using, e.g. shorter light pulses or wider tunable sources. The latter can be increased by using a different material with a stronger dispersion (that is, a lower Abbe number) than weakly dispersive fused silica glass or a narrower waveguide geometry with tighter transverse light confinement, e.g. on an integrated photonics platform~\cite{pelucchi2022potential}. This would increase the characteristic kinetic energy $\Delta \beta_{\rm char}'$ and would correspondingly allow for larger values of the coupling $C$ and shorter waveguide lengths.

\section{Observable signatures of the synthetic magnetic field}
\label{sec:observable}
\begin{figure}
	\includegraphics[width=\textwidth]{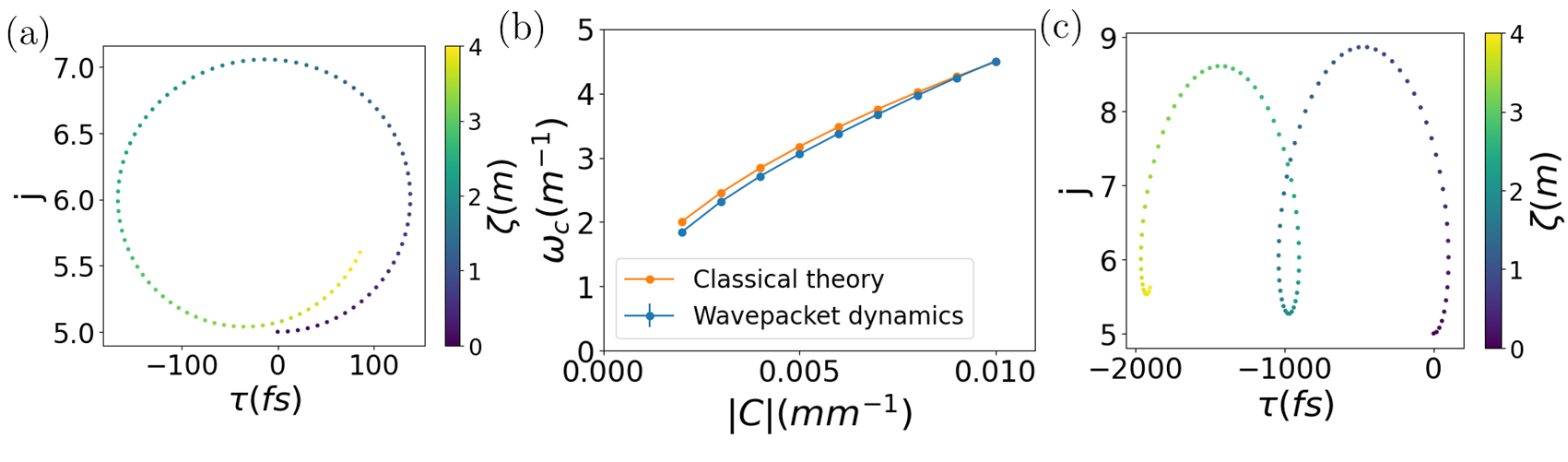}
	\centering
	\caption{(a): In our wavepacket dynamics with Eq.~\ref{eq:coupled_mode_cm} we find cyclotron orbits, as shown here by the wavepacket center-of-mass coloured according to the position along the optical axis, $\zeta$. (b): The frequencies of the cyclotron orbits extracted with a fit ({\it blue}) compared to the coupled wire model classical theory ({\it orange}), showing excellent agreement. (c): We apply a temperature gradient across the array, corresponding to an on-site potential. Tuning the strength of this potential introduces a Hall drift into the dynamics. The parameters used are the same as Fig.~\ref{fig:F2}. The example results in (a) and (c) use a coupling $C = -0.002$mm$^{-1}$. For the coupled-mode wavepacket dynamics, the wavepacket is prepared with an initial center-of-mass of $j_0 = 5$ and $\omega^{\prime} = 0$, and with widths of $s_{\omega^{\prime}} = 1 / 500 \times 10^{15}$Hz (corresponding to a Gaussian pulse length of 500fs), and $s_j = 1$. In (c) we use an electric potential of $\Delta n_j (\omega_0/c) = 0.001j$mm$^{-1}$.}
	\label{fig:F3}
\end{figure}
Having engineered our waveguide band structure in the co-moving frame to resemble that of a quantum Hall coupled wire model, we now numerically demonstrate novel optical effects that result from the synthetic magnetic field. These provide the smoking gun for non-trivial topological physics in this system.

\subsection{Cyclotron orbits}

As a first example, we consider the optical equivalent of bulk cyclotron orbits that arise for a semiclassical charged particle in a magnetic field. As discussed above, equispaced Landau levels are present in the bulk of the coupled wire model (Fig.~\ref{fig:F1}(b)). A wavepacket prepared in a suitable superposition of Landau levels will then execute semiclassical cyclotron orbits, moving in a circular trajectory with the characteristic cyclotron frequency set by the level spacing and a (clockwise or anti-clockwise) direction set by the sign of the effective magnetic field. In the presence of an additional synthetic electric field, the cyclotron motion will be supplemented by a so-called `Hall drift', that is a sideways motion perpendicular to the direction of the applied electric field.

As we now show, such orbits naturally arise for photons in our system. To this purpose, we prepare an initial Gaussian wavepacket in the $j-\omega^{\prime}$ space, spatially centered in the bulk of the array and with a central frequency located in the Landau level region of the bands. The Gaussian spatial width $s_j$ is taken of the order of the inter-waveguide spacing, while the chosen frequency-space width $s_{{\omega^\prime}}$ corresponds to a Gaussian pulse duration on the order of 100fs. Such pulse durations are well within the range of standard techniques in ultrafast optics such as mode-locked lasers, and the light then has to be focussed onto the input facet of the array with the appropriate spot waist to realise the desired Gaussian spatial profile. The wavepacket is then evolved along $\zeta$ according to the Fourier-space comoving-frame evolution equation (Eq.~\ref{eq:wave_omega_lab_frame}), and Fourier-transformed into $j-\tau$ space. The details of the numerical calculations throughout this section are discussed in the Supplemental Material~\cite{supmat}. 

Fig.~\ref{fig:F3}(a) shows an example trajectory of the pulse center-of-mass, calculated for the waveguide array parameters used in Fig.~\ref{fig:F2}. A clear cyclotron orbit is visible, where the amplitude of the oscillations along the spatial direction $j$ is of the order of a waveguide, so they are detectable in experiment. In order to further characterize the oscillations, we repeat these simulations for different values of the inter-waveguide coupling $C$, which physically corresponds to varying their spacing. For each calculation, the cyclotron frequency is extracted from a fit of the $\zeta$-evolution, the details of which are discussed in the Supplemental Material~\cite{supmat}. The blue line in Fig.~\ref{fig:F3}(b) shows the value of the fitted cyclotron frequency as a function of the coupling $C$; as expected, it grows for increasing values of the coupling. 

A deeper insight into the cyclotron oscillations can be obtained by comparing these numerical results with the prediction of an approximate classical calculation based on the equations of motion for a classical particle with constant, yet anisotropic masses and subject to the synthetic magnetic field according to the coupled wire model with no external potential~\cite{Ozawa2017}. This calculation leads to the prediction
\begin{equation}
    \omega_c = \frac{|\mathcal{B}|}{\sqrt{m^{(\tau)}\,m^{(j)}}} = |\mathcal{B}|\,\sqrt{2\,\left|C\,D_{j_{\rm ref}}\right|}
\end{equation}
where $|m^{(\tau),(j)}|=1/|D_{j_{\rm ref}}|,1/(2\,|C|)$ are the absolute values of the effective masses in the temporal and spatial directions respectively~\footnote{Depending on the relative sign of the masses in the two directions (namely of $C$ and $D_{j_{\rm ref}}$), the Landau levels appear for states displaying the same or opposite phases in neighboring wires.}, and $\mathcal{B} \equiv (A^{(\tau)}_{j = N} - A^{(\tau)}_{j = 1}) / (N - 1)$ is the approximately uniform magnetic field corresponding to our magnetic vector potential. The result of this approximate calculation is shown as an orange line in Fig.~\ref{fig:F3}(b) and displays a good agreement with the numerics for the full model ({\it blue}). The small deviation between the two curves is principally due to the minor differences between the ideal coupled-wire model bands and our full optical results in Fig.~\ref{fig:F2}. 

For the chosen system parameters, the typical period $\zeta_c = 2\pi / \omega_c$ of the cyclotron oscillations is of the order of metres, which may be very demanding compared to the typical length of waveguide arrays fabricated in experiment. However, this difficulty could be mitigated by using state recycling techniques to increase the effective lengthscale explored in experiment~\cite{Mukherjee2018b} or by switching to different material platforms. The cyclotron orbit length-scale is in fact determined by the characteristic kinetic energy scale $\Delta\beta'_{\rm char}$ which, as discussed above, may be increased using a wider operating bandwidth or samples with a stronger dispersion.

\subsection{Hall drift on the pulse arrival time}

We now exploit another feature of quantum Hall physics to introduce a Hall drift into the cyclotron dynamics we found above. As mentioned previously, if an additional electric field is applied to a particle in a quantum Hall system, we expect the particle to drift in the in-plane direction perpendicular to that field. We first consider applying a synthetic electric field across the array in the $j$-direction, which will correspond to a drift in the $\tau$-direction. If this drift could be controlled, natural applications of the resulting delay/advance include delay lines, which have a widespread uses throughout optics, including interferometry, ultrafast optics and telecommunications.

The Hall drift can be introduced, with controllable magnitude and direction, by imposing suitable perturbations to the waveguide array, e.g. a temperature gradient along $j$. This induces a corresponding spatial gradient  of the refractive index $\Delta n_j$ along $j$. Formally, this can be described by including an additional term of the form:
\begin{equation}
    i\frac{\partial \tilde{a}^\prime_j}{\partial \zeta}= \ldots +\Delta n_j (\omega_0/c)\,\tilde{a}^\prime_j
\end{equation}
to the right-hand side of our model equation (Eq.~\ref{eq:wave_omega_lab_frame}). An elementary calculation within the coupled wire model shows that the temporal drift under a synthetic electric field $\mathcal{E}_{\text{therm}} = -(\omega_0/c) (d\Delta n_j/dj)$ in the spatial direction is equal to
\begin{equation}
    \tau_{H}=-\zeta\,\frac{\mathcal{E}_{\text{therm}}}{\mathcal{B}}\,.
\end{equation}
An example of this effect is illustrated in Fig.~\ref{fig:F3}(c), where we show the appearance of the Hall drift along $\tau$ under the effect of a potential gradient as small as $\Delta n_j (\omega_0/c) = 0.001\,j\,\textrm{mm}^{-1}$, corresponding to a refractive index perturbation of $\sim 10^{-7}$. We note that the upper limit on the perturbation strength is due to the size of the band gap in Fig.~\ref{fig:F2}(c); a perturbation of the order of or larger than the gap will introduce significant non-adiabatic effects and blur out the effect shown in Fig.~\ref{fig:F3}(c).

\subsection{Hall drift in real space}
\begin{figure*}[t]
	\includegraphics[width=\textwidth]{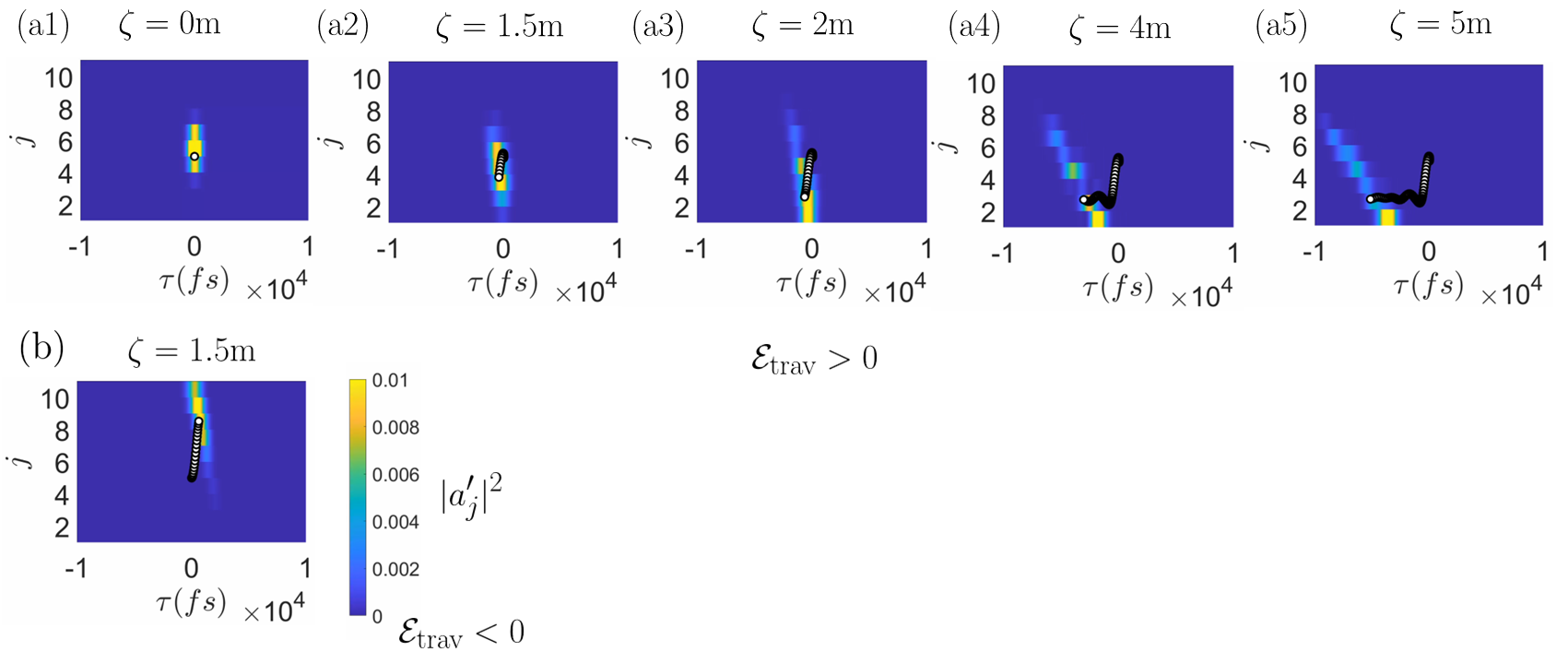}
	\centering
	\caption{(a1) - (a5): We apply a travelling refractive index modulation to create an effective electric field in the $\tau$ direction, which causes a displacement of the wavepacket across the waveguides as we expect for a quantum Hall system. We use the same system parameters as Fig.~\ref{fig:F3}, with an additional electric field $\mathcal{E}_{\text{trav}} = 0.00001 $(fs mm)$^{-1}$. (b): Reversing the sign of the refractive index modulation reverses the electric field to $\mathcal{E}_{\text{trav}} = -0.00001 $(fs mm)$^{-1}$ and, thus, the direction of displacement.}
	\label{fig:F4}
\end{figure*}
In the previous Subsection, we showed how the displacement along the $\tau$-direction can be introduced. Now we propose a method to exploit the same quantum Hall effect to control the spatial displacement of a wavepacket across the array, i.e. along the $j$-direction, in response to a perturbation along the temporal $\tau$ direction.

The idea of the scheme is to implement a spatial displacement in a reconfigurable way by means of a travelling refractive index perturbation. This could be realised in experiments by means of the electro-optic or acousto-optical effects, as was recently investigated in~\cite{wang_multidimensional_2020,Piccioli2021}. In the simplest case, we consider a refractive index perturbation that is uniform across the array and travels along the waveguides at the same speed as the reference group velocity $v_g^{\rm (ref)}$:
\begin{equation}
    \Delta n(z,t)=\Delta\bar{n}\,\frac{z - v_g^{\rm (ref)}t}{\ell},
\end{equation}
where $\ell$ is the length of the device in $z$ and $\Delta\bar{n}$ is the strength of the perturbation. Such a perturbation can be included in our model by adding a term of the form:
\begin{equation}
    i\frac{\partial a^\prime_j}{\partial \zeta}= \ldots - \mathcal{E}_{\text{trav}}\tau a_j^{\prime}
\end{equation}
to the right-hand side of the Schr\"{o}dinger equation (Eq.~\ref{eq:schrodinger}), where $\mathcal{E}_{\text{trav}} = \Delta\bar{n}\,\omega_0\,v_g^{\rm (ref)}/(\ell c)$ is the synthetic electric field along the $\tau$-direction. Two examples of light propagation under such a perturbation are shown in Fig.~\ref{fig:F4}(a1) - (a2) and (b) for the same magnitudes of the synthetic electric field but opposite signs. We see the wavepacket, prepared in the system bulk, transported across the array towards larger or smaller $j$. These numerics are carried out using the frequency space Eq.~\ref{eq:coupled_mode_cm}, where the perturbation appears as a term of the form $i\mathcal{E}_{\text{trav}}\partial \tilde{a}_j^{\prime} / \partial\omega^{\prime}$ added to the right-hand side. Note also that the displacement appears in spite of the modulation being independent of $j$; this further confirms its origin from the synthetic magnetic field via the quantum Hall effect.

As an alternative scheme, the same effect could be realized by means of a variation of the magnetic vector potential $A^{(\tau)}_j$ in the Schr\"{o}dinger equation (Eq.~\ref{eq:schrodinger}) during the evolution. According to the $t-z$ mapping, this requires us to vary the waveguide properties along the waveguide axis. Analogously to classical electrodynamics, this produces an effective synthetic electric field $\mathcal{E}_{\text{z-mod}} = - dA_{j}^{(\tau)}(\zeta) / d\zeta$ oriented along the $\tau$ direction which, by the quantum Hall effect, induces a drift along the orthogonal spatial direction $j$. As an example, the spatial gradient of $A^{(\tau)}_j(z)$ in the $z$-direction could be obtained in our setup by designing the waveguide parameters as:
\begin{eqnarray}
    \delta n_j(z)&=&\delta n_{j-w z}(z=0) \\
    \sigma_j(z)&=&\sigma_{j-wz}(z=0)
\end{eqnarray}
where evaluation of the waveguide parameters at the continuous-valued $j - wz$ is obtained by interpolating their values at $z = 0$ in-between neighboring waveguides. Within the coupled wire model, this leads to
\begin{equation}
    \mathcal{E}_{\text{z-mod}} = - \frac{dA_{j}^{(\tau)}(\zeta)}{d\zeta} = -w\mathcal{B}.
\end{equation}
and the corresponding Hall drift in the spatial direction can be straightforwardly evaluated to be:
\begin{equation}
    j_H = -\zeta\,\frac{\mathcal{E}_{\text{z-mod}}}{\mathcal{B}}
\end{equation}
In spite of their different optical implementation, it is worth noting that these two approaches are actually the same from the point of view of the band structure. Both of them are in fact based on an adiabatic transport of the state along the band, corresponding to a change in the spatial position along the $j$-direction, as indicated by the colouring of the bands in Fig.~\ref{fig:F2}(c).

\subsection{Propagation along the edge}

The plots in Fig.~\ref{fig:F4}(a1), (a2) and (b) refer to relatively short propagation distances, so that the Hall-drifted wavepackets are still within the bulk of the waveguide array. At longer propagation distances, the wavepacket can reach the spatial edge of the waveguide array at $j = 1$ or $j = N$. At this point, as is usual in topological systems under a synthetic electric field~\cite{DeBernardis2023}, the wavepacket gets converted into an edge excitation which propagates along the edge. The ensuing fast chiral motion along the spatial edge of the system is clearly visible in the plots in Fig.~\ref{fig:F4}(a3) - (a5), as well as in the animation that is available as Supplemental Material~\cite{supmat}; since we are dealing with a spatial edge, the chiral motion is along the temporal $\tau$ direction, with a different sign depending on which edge the wavepacket hits. Finally, we note that we do not have an edge in the $\tau$-direction, so any advance or delay that we see from either the temperature gradient or from this chiral edge mode propagation could be of arbitrary size. 

\section{Conclusions}
\label{sec:conclu}
In this work, we have demonstrated a novel framework to generate a synthetic magnetic field for light in a one-dimensional array of coupled waveguides. Based on the $t-z$ mapping of paraxial propagation of time-dependent optical pulses, a two-dimensional model is obtained in the $j-t$ plane spanned by the (discrete) waveguide index and the (continuous) temporal variable. With a suitable gradient of the waveguide properties across the array, an effective synthetic magnetic field is induced which provides an optical realization of the quantum Hall coupled wire model. Observable signatures of the synthetic magnetic effects are anticipated as a chiral cyclotron motion in the $j-t$ plane, a Hall drift in the temporal or spatial direction under the effect of a synthetic electric field directed along the array or in the temporal direction, and a fast propagation in chiral edge states. 

From an experimental point of view, important advantages of our proposal over previous work on synthetic dimensions in photonics~\cite{yuan2018synthetic,Ozawa2019} can be pointed out. 
Building atop available schemes for topological photonics in waveguide arrays~\cite{Rechtsman2013a,Ozawa2019}, our proposal only requires working with time-dependent pulses rather than monochromatic light and, in particular, it does not require any external dynamical modulation of the system and does not involve all the complexities of Floquet systems. Eventually, it will open new avenues for the spatio-temporal manipulation of optical pulses~\cite{Ozawa2021, Nemirovsky2021}. 

Generalization of our proposal to physically two-dimensional waveguide arrays suggests a natural way to realize three-dimensional models: future work will be devoted to the investigation of topological models involving two discrete $j_{1,2}$ coordinates and a continuous $t$ one in our platform and the identification of observable optical signatures of the geometrical and topological properties of the peculiar features of three-dimensional band structures such as Weyl points and Fermi arcs.

While this work was focused on a specific implementation of our proposed concept in laser-written waveguide array operating in the visible light domain, future work will be devoted to the identification of alternative realizations in different material systems and frequency domains, e.g. integrated photonics devices for infrared/visible light~\cite{pelucchi2022potential} or microwave waveguides~\cite{esposito2021perspective}, which may provide a more pronounced dispersion as well as strong nonlinearities.

In the long run, our proposal holds in fact great promise in view of realizing novel states of topological photonic matter. 
In contrast to topological models exploiting the light frequency as a synthetic dimension~\cite{Ozawa2016, Yuan2016,dutt_creating_2022,wang_multidimensional_2020}
where nonlinearities would typically result in long-range interactions along the frequency direction~\cite{Ozawa2017}, the fact that the spatial coordinates of the topological model are encoded in the waveguide index $j$ and the temporal variable $t$ translates a spatially local nonlinearity of the medium into local interactions in the topological model. This feature is of extreme importance~\cite{ozawa2019topological} when one is to scale up the interaction strength and realize strongly correlated states like fractional quantum Hall liquids of light~\cite{carusotto2013quantum,carusotto2020photonic}.

\section{Acknowledgements}
I.C. acknowledges financial support from the Provincia Autonoma di Trento, from the Q@TN initiative, and from PNRR MUR project PE0000023-NQSTI. C.O. and H. M. P. are supported by the Royal Society via grants UF160112, RGF/EA/180121 and RGF/R1/180071. M.C.R. gratefully acknowledges support from the Office of Naval Research under agreement number N00014-23-1-2102 as well as the Air Force Office of Scientific Research MURI program under agreement number FA9550-22-1-0339. S.M. gratefully acknowledges support from IISc and SERB (SRG/2022/002062). Illuminating discussions with Daniele De Bernardis and Francesco Piccioli are warmly acknowledged.

\clearpage
\begin{center}
\textbf{\large Supplemental Material for\\ ``Artificial gauge fields in the t-z mapping for optical pulses: spatio-temporal wavepacket control and quantum Hall physics"}
\end{center}
\setcounter{equation}{0}
\setcounter{figure}{0}
\setcounter{table}{0}
\setcounter{page}{1}
\setcounter{section}{0}
\makeatletter
\renewcommand{\theequation}{S\arabic{equation}}
\renewcommand{\thefigure}{S\arabic{figure}}
\renewcommand{\bibnumfmt}[1]{[S#1]}
\renewcommand{\citenumfont}[1]{S#1}
\renewcommand{\thesection}{S\arabic{section}}

We cover three topics in this Supplemental Material. In Sec.~\ref{sec:analytic_waveguides}, we discuss a simple, analytically-solvable toy model for a 1D coupled waveguide array, which we use to demonstrate that the idea of engineering the co-moving frame band structure to have quantum Hall features is quite general and not specific to the particular model chosen. For the other two sections, we return to the experimentally-motivated model introduced in the Main Text. In Sec.~\ref{sec:schrodinger_quants}, we show the calculation of the Schr\"{o}dinger equation effective mass, magnetic vector potential and on-site potential which we use in some of the results in the Main Text. Finally, in Sec.~\ref{sec:wpd}, we give the technical details of the wavepacket dynamics numerical simulations and data analysis that are employed in the Main Text.

\section{An Analytical Toy Model for Coupled Waveguides}
\label{sec:analytic_waveguides}

To further demonstrate the engineering of non-trivial magnetic field effects from tuning waveguide parameters as discussed in the Main Text, we now consider a simple, analytical model for a coupled waveguide array, consisting of metal waveguides each embedding a medium of frequency-independent refractive index $n_j$. We can exactly solve the paraxial Helmholtz equation to find that the TE$_{10}$ modes, which we choose for simplicity, have the dispersion~\cite{Grant2004}:
\begin{equation}
    \beta_j(\omega) = \frac{n_j}{c}\sqrt{\omega^2 - \omega_c^{(j)}{}^2},
    \label{eq:analytic_pcs}
\end{equation}
where $\omega_c^{(j)} = \pi c / (L_x^{(j)}n_j)$ is the cutoff frequency in waveguide $j$, below which no modes can propagate. The cutoff frequency depends on the waveguide width $L_x^{(j)}$ which we allow one to spatially vary in order to engineer a non-trivial magnetic field.
\begin{figure}[H]
	\includegraphics[width=\textwidth]{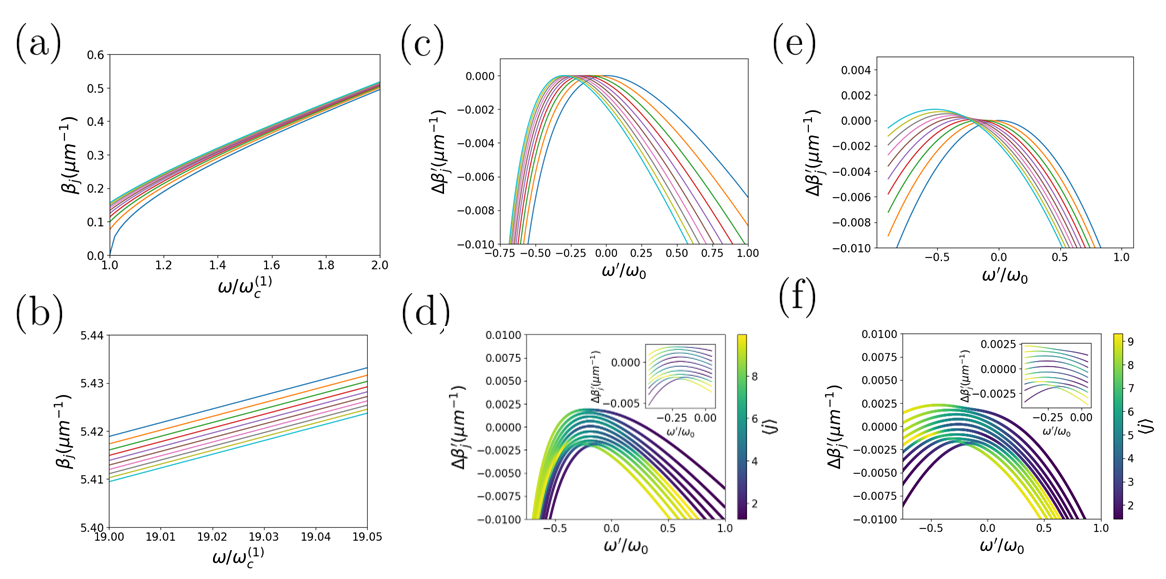}
	\centering
	\caption{(a): Propagation constants in the lab frame for our toy model. The blue labelled curve is the dispersion for waveguide $j = 1$. (b): The dispersion curves in (a) for larger frequencies, showing that the order of the curves has reversed and hence that the dispersions cross over each other. (c): Direct mapping of the dispersions in (a) to the co-moving frame using our modified Doppler shift. We see the maxima of the dispersion curves all have the same $\Delta\beta_j^{\prime}$ value as we expect. (d): Results in (c) now including a coupling between neighbouring waveguides, showing avoided crossings. This results in a band structure resembling the coupled wire model, including bulk Landau levels (c.f. inset) and chiral edge modes in the gap. (e): Propagation constants in the co-moving frame for this model calculated using our Sch\"{o}dinger equation without any coupling, showing qualitative agreement with our analytical results in (c). (f): Results in (e) including a coupling between neighbouring waveguides. Throughout, we use $N = 10$ waveguides with $L_0 = 10\mu$m, $\Delta L = 1\mu$m and $n_1 = 1.5$, and we take $\omega_0 = 10\omega_c^{(1)}$. We take the coupling to be $C = 10^3\mu$m$^{-1}$.}
	\label{fig:S2}
\end{figure}
We can now transform the dispersion into the co-moving frame using the mapping in the Main Text (Eq. 10). Note that, in the co-moving frame, we actually plot $\Delta\beta^{\prime}_j(\omega^{\prime}) = \beta^{\prime}_j(\omega^{\prime}) - \beta^{\prime}_{\text{ref}}(0)$ for consistency with our definition of propagation constants in our Schr\"{o}dinger equation. In our calculations here, we choose the reference waveguide to be the $j = 1$ waveguide in the array and we take $\omega_0 = 10\,\omega_c^{(1)}$ as the carrier frequency. 

We can then use the transformed dispersions to tune the refractive index $n_j$ such that the maxima of the dispersions in the co-moving frame all take the same value, which makes the resultant band structure as similar as possible to the coupled wire model (discussed in the Main Text). We tune our refractive index profile as:
\begin{equation}
    n_j = \frac{n_1}{\sqrt{1 + \left(\frac{\omega_c^{(1)}}{\omega_0}\right)^2\left(\left(\frac{L_x^{(j)}}{L_x^{(1)}}\right)^2 - 1\right)}}
\end{equation}
which we calculate by differentiating our transformed propagation constants with respect to $\omega^{\prime}$ and enforcing that the maxima are all equal in $\Delta\beta_j^{\prime}$. We also choose the waveguide widths $L_x^{(j)} = L_0 + \Delta L\sqrt{j}$ to ensure that the spacing between adjacent dispersion curves is approximately constant, to approach the case of a uniform magnetic field. The resulting dispersions in both frames are shown in Fig.~\ref{fig:S2}(a), (b) and (c). In the lab frame, we see a set of dispersion curves that cross over each other (c.f. the reversed order of the curves in (a) vs. (b)). In the co-moving frame, we have a set of curves with maxima that are all at the same $\Delta\beta_j^{\prime}$ value as expected from our chosen refractive index profile (c.f. panel (c)). Introducing a coupling between neighbouring waveguides results in avoided crossings (Fig.~\ref{fig:S2}(d)), and we see a band structure that resembles that of the quantum Hall coupled wire model (c.f. Fig. 1(b) in the Main Text). In particular, in the inset of the figure, we see flat Landau level states in the bulk and chiral edge states in the gap, which is characteristic of quantum Hall systems.  Furthermore, panel (a) suggests that a stronger dispersion could be obtained by working at a lower value of $\omega_0 / \omega_c^{(1)}$, which corresponds to a tighter waveguide confinement, that is, a smaller $L_0$.

We can also calculate the band structure from our Schr\"{o}dinger equation (Eq. 6 in the Main Text) by using our propagation constants (Eq.~\ref{eq:analytic_pcs}) to calculate $m_j, A^{(\tau)}_j$ and $V_j$, and then diagonalising the resulting Hamiltonian. Example results are shown in Fig~\ref{fig:S2}(e) and (f), without and with an inter-waveguide coupling respectively. Comparing to the corresponding exact results in (c) and (d), we see that, as expected, the Schr\"{o}dinger equation captures the dispersions very well close to $\omega^{\prime} = 0$ because the truncated Taylor expansion is most accurate there. Moving away from $\omega^{\prime} = 0$ in either direction leads to disagreement between the two approaches, most notably in the heights of the maxima not being identical in (e), leading to dispersion even in the flat Landau level states in (f). Finally, we note that the bands we find for this model are inverted relative to the bands for the experimentally-motivated model we consider in the Main Text (i.e. the bands for the toy model have stationary points that are maxima, not minima). This is because the group velocity dispersions in the two models have opposite signs. Overall, these results therefore demonstrate that we do not require a complex model to engineer the kind of physics we find here.

\section{Schr\"{o}dinger equation effective mass, magnetic vector potential and on-site potential}
\label{sec:schrodinger_quants}
\begin{figure*}
	\includegraphics[width=\textwidth]{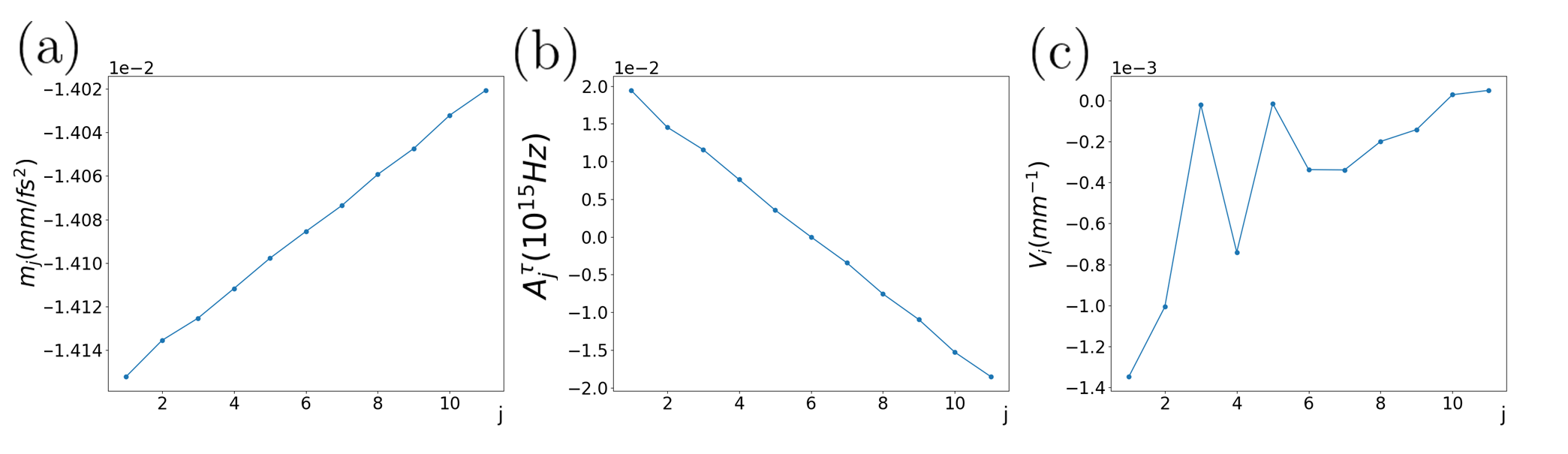}
	\centering
	\caption{Effective mass, magnetic vector potential and on-site potential in panels (a), (b) and (c) respectively for our Schr\"{o}dinger equation, calculated from data in Fig. 3 in the Main Text. We see a weakly-varying effective mass, near-linear magnetic vector potential (corresponding to a uniform magnetic field) and a small residual on-site potential.}
	\label{fig:schro_quant}
\end{figure*}
In this section, we show the results of using the waveguide dispersions calculated in the Main Text to find the effective mass, magnetic vector potential and on-site potential in the Schr\"{o}dinger equation, which we then use to calculate some of the results in Fig. 3. We use the lab-frame propagation constants (Fig. 3(a)) and evaluate the three quantities using Eq. 7 in the Main Text. The results are shown in Fig.~\ref{fig:schro_quant}. We see a near-constant effective mass (with variation on the order 1\% across the array), a magnetic vector potential that is very close to linear in $j$ (corresponding to a uniform magnetic field), and a very small on-site potential.  

\section{Details of Wavepacket Dynamics Simulations}
\label{sec:wpd}
\begin{figure*}
	\includegraphics[width=\textwidth]{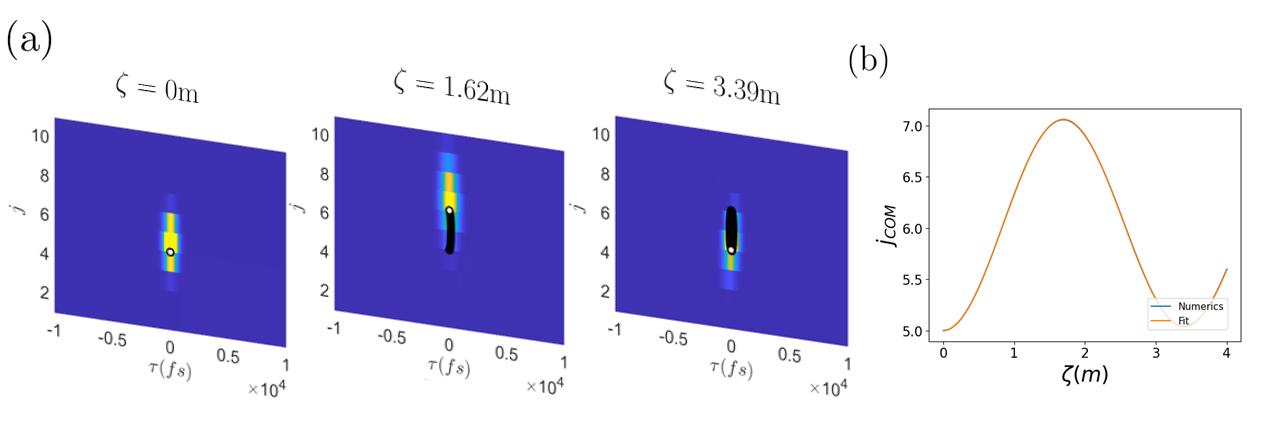}
	\centering
	\caption{(a): Examples of wavepacket density plotted at different evolution times $\zeta$ starting from a Gaussian wavepacket prepared in the bulk. We see part of a cyclotron orbit with a Hall drift. (b): Plot of the center-of-mass of the wavepacket in (a) in the discrete direction as a function of $\zeta$ together with a fit to Eq.~\ref{eq:fit}, showing that the fitting function captures our data very well, and can be used to extract the cyclotron frequency. We use the same parameters as the wavepacket dynamics simulations in Fig. 4 in the Main Text, with a coupling $C = -0.002$mm$^{-1}$.}
	\label{fig:S3}
\end{figure*}
As discussed in the Main Text, we use wavepacket dynamics simulations to investigate bulk cyclotron orbits and other physical observables. We prepare a Gaussian wavepacket of the form:
\begin{equation}
    \tilde{a}_j^{\prime}(\zeta = 0, \omega^{\prime}) = Ae^{ik_j(j - j_0)}e^{-\frac{(j - j_0)^2}{2s_j^2}} e^{ik_{\omega^{\prime}}({\omega^{\prime}} - {\omega^{\prime}}_0)}e^{-\frac{({\omega^{\prime}} - {\omega^{\prime}}_0)^2}{2s_{\omega^{\prime}}^2}},
\end{equation}
where $A$ is a normalisation constant; $k_{(j, \omega^{\prime})}$ is the wavepacket momentum in the two directions; $(j,\omega^{\prime})_0$ is the initial center-of-mass position and $s_{(j, \omega^{\prime})}$ is the wavepacket width along the two directions. We choose the wavepacket to be localised in the $j$-bulk and to have an initial frequency in the bulk part of the bands in order to target Landau level states. We then numerically propagate the wavepacket through the array by discretising the $\omega^{\prime}$ dimension into $M \gg 1$ points and hence representing our initial wavepacket as an $NM$-component column vector $\underline{a}(\zeta = 0)$. We then evolve the initial vector via:
\begin{equation}
    \underline{a}(\zeta + \delta\zeta) = e^{-iH\delta\zeta}\underline{a}(\zeta),
\end{equation}
where $\delta\zeta$ is our small `timestep' and $H$ is the $NM \times NM$ matrix representing the right-hand side of the coupled mode equation (Eq. 8 in the Main Text) in our finite difference basis. More precisely, we have:
\begin{equation}
    H = H_{\text{diag}} - 
    \begin{pmatrix}
        0 & C & 0 & \cdots & 0 \\
        C & \ddots & \ddots & \ddots & \vdots \\
        0 & \ddots & & & 0 \\
        \vdots & \ddots & & & C \\
        0 & \cdots & 0 & C & 0
    \end{pmatrix} \otimes I_{\omega^{\prime}},
\end{equation}
where $H_{\text{diag}}$ is a diagonal matrix formed by placing (-1 times) the discretised co-moving frame propagation constants $\beta_j(\omega) - \beta_{j_{\text{ref}}}(\omega_0) - \omega^{\prime} / v_g^{\text{(ref)}}$ along the main diagonal. The second term, representing the coupling between neighbouring waveguides, is an $N \times N$ tridiagonal matrix with the couplings inserted on to the two diagonals either side of the main diagonal. We then Fourier transform the state with respect to $\omega^{\prime}$ to map it into $\tau$-space, and consider the corresponding $j-\tau$ density as a function of $\zeta$. An example of the wavepacket density for the optical model is shown in Fig.~\ref{fig:S3}(a), corresponding to the center-of-mass trajectory plotted in Fig.~4(a) of the Main Text. We use this density to calculate the wavepacket center-of-mass as a function of $\zeta$, $j_{\text{COM}}(\zeta)$ and $\tau_{\text{COM}}(\zeta)$.

When investigating bulk cyclotron orbits, we fit to the $j_{\text{COM}}(\zeta)$ trajectory using the function:
\begin{equation}
    f(\zeta) = A\sin(\omega_c\zeta + \phi)e^{-g\zeta} + B,
    \label{eq:fit}
\end{equation}
and extract the cyclotron frequency $\omega_c$, an example of which is shown in Fig.~\ref{fig:S3}(b). We choose this function because we expect cyclotron orbits to be circular trajectories in $(j, \tau)$. We include an exponential damping factor to take into account the small damping seen in some of our numerics. We see that the fit captures our numerical data well, and it performs similarly well on all our numerics.

As discussed in the Main Text, besides the bulk cyclotron orbits, we also investigate two other physical observables. The first is a travelling refractive index perturbation. We model this with a term of the form $i\mathcal{E}_{\text{trav}}\partial\tilde{a}_j^{\prime} / \partial\omega^{\prime}$ which corresponds, in the Schr\"{o}dinger equation, to an effective electric field along the $\tau$ direction, $-\mathcal{E}_{\text{trav}}\tau a_j^{\prime}$. We include this term within our numerical scheme by representing the $\partial/\partial\omega^{\prime}$ operator by a standard $M \times M$ finite difference first derivative matrix $d_{\omega^{\prime}}$. We hence include the term in our total finite difference matrix as $H \to H + i\mathcal{E}_{\text{trav}}I_j\otimes d_{\omega^{\prime}}$, where $I_j$ is the $N \times N$ identity matrix. Finally, we also consider applying a temperature gradient across the array, which we model with a term of the form $\Delta n_j(\omega_0 / c)\tilde{a}_j^{\prime}$, where we choose $\Delta n_j = Uj$. We include this in our numerical scheme by $H \to H + UJ\otimes I_{\omega^{\prime}}$, where $J = \text{diag}(1,...,N)$ and $I_{\omega^{\prime}}$ is the $M \times M$ identity matrix.


\begin{thebibliography}{10}
\newcommand{\enquote}[1]{``#1''}

\bibitem{Boyd2008}
R.~W. Boyd, \emph{Nonlinear optics} (Academic Press, London, 2008), 3rd ed.

\bibitem{Szameit2010}
A.~Szameit and S.~Nolte, \enquote{Discrete optics in femtosecond-laser-written
  photonic structures,} {{Journal of Physics B: Atomic,
  Molecular and Optical Physics}} \textbf{43}, 163001 (2010).

\bibitem{Schwartz2007}
T.~Schwartz, G.~Bartal, S.~Fishman, and M.~Segev, \enquote{Transport and
  {A}nderson localization in disordered two-dimensional photonic lattices,}
  {{Nature}} \textbf{446}, 52 -- 55 (2007).

\bibitem{Levi2011}
L.~Levi, M.~Rechtsman, B.~Freedman, T.~Schwartz, O.~Manela, and M.~Segev,
  \enquote{Disorder-enhanced transport in photonic quasicrystals,}
  {{Science}} \textbf{332}, 1541--1544 (2011).

\bibitem{Verbin2013}
M.~Verbin, O.~Zilberberg, Y.~E. Kraus, Y.~Lahini, and Y.~Silberberg,
  \enquote{Observation of topological phase transitions in photonic
  quasicrystals,} {{Phys. Rev. Lett.}} \textbf{110},
  076403 (2013).

\bibitem{Rechtsman2013a}
M.~C. Rechtsman, J.~M. Zeuner, A.~Tünnermann, S.~Nolte, M.~Segev, and
  A.~Szameit, \enquote{Strain-induced pseudomagnetic field and photonic
  {L}andau levels in dielectric structures,} {{Nature
  Photon}} \textbf{7}, 153–158 (2013).

\bibitem{Mukherjee2018}
S.~Mukherjee, M.~Di~Liberto, P.~\"{O}hberg, R.~R. Thomson, and N.~Goldman,
  \enquote{Experimental observation of {A}haronov-{B}ohm cages in photonic
  lattices,} {{Phys. Rev. Lett.}} \textbf{121}, 075502
  (2018).

\bibitem{Szameit2011}
A.~Szameit, M.~C. Rechtsman, O.~Bahat-Treidel, and M.~Segev,
  \enquote{$\mathcal{P}\mathcal{T}$-symmetry in honeycomb photonic lattices,}
  {{Phys. Rev. A}} \textbf{84}, 021806 (2011).

\bibitem{Segev1992}
M.~Segev, B.~Crosignani, A.~Yariv, and B.~Fischer, \enquote{Spatial solitons in
  photorefractive media,} {{Phys. Rev. Lett.}}
  \textbf{68}, 923--926 (1992).

\bibitem{Eisenberg1998}
H.~S. Eisenberg, Y.~Silberberg, R.~Morandotti, A.~R. Boyd, and J.~S. Aitchison,
  \enquote{Discrete spatial optical solitons in waveguide arrays,}
  {{Phys. Rev. Lett.}} \textbf{81}, 3383--3386 (1998).

\bibitem{Efremidis2002}
N.~K. Efremidis, S.~Sears, D.~N. Christodoulides, J.~W. Fleischer, and
  M.~Segev, \enquote{Discrete solitons in photorefractive optically induced
  photonic lattices,} {{Phys. Rev. E}} \textbf{66}, 046602
  (2002).

\bibitem{Christodoulides2003}
D.~Christodoulides, F.~Lederer, and Y.~Silberberg, \enquote{Discretizing light
  behaviour in linear and nonlinear waveguide lattices,}
  {{Nature}} \textbf{424}, 817 -- 823 (2003).

\bibitem{Fleischer2003}
J.~W. Fleischer, M.~Segev, N.~K. Efremidis, and D.~N. Christodoulides,
  \enquote{Observation of two-dimensional discrete solitons in optically
  induced nonlinear photonic lattices,} {{Nature}}
  \textbf{422}, 147 -- 150 (2003).

\bibitem{Lahini2008}
Y.~Lahini, A.~Avidan, F.~Pozzi, M.~Sorel, R.~Morandotti, D.~N. Christodoulides,
  and Y.~Silberberg, \enquote{Anderson localization and nonlinearity in
  one-dimensional disordered photonic lattices,} {{Phys.
  Rev. Lett.}} \textbf{100}, 013906 (2008).

\bibitem{Freedman2006}
B.~Freedman, G.~Bartal, M.~Segev, R.~Lifshitz, D.~N. Christodoulides, and J.~W.
  Fleischer, \enquote{Wave and defect dynamics in nonlinear photonic
  quasicrystals,} {{Nature}} \textbf{440}, 1166–1169
  (2006).

\bibitem{Larre2015a}
P.-E. Larr\'e and I.~Carusotto, \enquote{Optomechanical signature of a
  frictionless flow of superfluid light,} {{Phys. Rev. A}}
  \textbf{91}, 053809 (2015).

\bibitem{fontaine2020interferences}
Q.~Fontaine, P.-{\'E}. Larr{\'e}, G.~Lerario, T.~Bienaim{\'e}, S.~Pigeon,
  D.~Faccio, I.~Carusotto, {\'E}.~Giacobino, A.~Bramati, and Q.~Glorieux,
  \enquote{Interferences between {B}ogoliubov excitations in superfluids of
  light,} {{Physical Review Research}} \textbf{2}, 043297
  (2020).

\bibitem{braidotti2022measurement}
M.~C. Braidotti, R.~Prizia, C.~Maitland, F.~Marino, A.~Prain, I.~Starshynov,
  N.~Westerberg, E.~M. Wright, and D.~Faccio, \enquote{Measurement of {P}enrose
  superradiance in a photon superfluid,} {{Physical Review
  Letters}} \textbf{128}, 013901 (2022).

\bibitem{steinhauer2022analogue}
J.~Steinhauer, M.~Abuzarli, T.~Aladjidi, T.~Bienaim{\'e}, C.~Piekarski, W.~Liu,
  E.~Giacobino, A.~Bramati, and Q.~Glorieux, \enquote{Analogue cosmological
  particle creation in an ultracold quantum fluid of light,}
  {{Nature Communications}} \textbf{13}, 2890 (2022).

\bibitem{Ozawa2019}
T.~Ozawa, H.~M. Price, A.~Amo, N.~Goldman, M.~Hafezi, L.~Lu, M.~C. Rechtsman,
  D.~Schuster, J.~Simon, O.~Zilberberg, and I.~Carusotto, \enquote{Topological
  photonics,} {{Rev. Mod. Phys.}} \textbf{91}, 015006
  (2019).

\bibitem{price2022roadmap}
H.~Price, Y.~Chong, A.~Khanikaev, H.~Schomerus, L.~J. Maczewsky, M.~Kremer,
  M.~Heinrich, A.~Szameit, O.~Zilberberg, Y.~Yang \emph{et~al.},
  \enquote{Roadmap on topological photonics,} {{Journal of
  Physics: Photonics}} \textbf{4}, 032501 (2022).

\bibitem{Rechtsman2013b}
M.~C. Rechtsman, J.~M. Zeuner, Y.~Plotnik, Y.~Lumer, D.~Podolsky, F.~Dreisow,
  S.~Nolte, M.~Segev, and A.~Szameit, \enquote{Photonic {F}loquet topological
  insulators,} {{Nature}} \textbf{496}, 196–200 (2013).

\bibitem{lindner2011floquet}
N.~H. Lindner, G.~Refael, and V.~Galitski, \enquote{Floquet topological
  insulator in semiconductor quantum wells,} {{Nature
  Physics}} \textbf{7}, 490--495 (2011).

\bibitem{Lumer2013}
Y.~Lumer, Y.~Plotnik, M.~C. Rechtsman, and M.~Segev, \enquote{Self-localized
  states in photonic topological insulators,} {{Phys. Rev.
  Lett.}} \textbf{111}, 243905 (2013).

\bibitem{Mukherjee2021}
S.~Mukherjee and M.~C. Rechtsman, \enquote{Observation of unidirectional
  solitonlike edge states in nonlinear {F}loquet topological insulators,}
  {{Phys. Rev. X}} \textbf{11}, 041057 (2021).

\bibitem{Mukherjee2020}
S.~Mukherjee and M.~C. Rechtsman, \enquote{Observation of {F}loquet solitons in
  a topological bandgap,} {{Science}} \textbf{368},
  856--859 (2020).

\bibitem{Noh2017}
J.~Noh, S.~Huang, D.~Leykam, Y.~D. Chong, K.~P. Chen, and M.~C. Rechtsman,
  \enquote{Experimental observation of optical {W}eyl points and {F}ermi
  arc-like surface states,} {{Nature Phys}} \textbf{13},
  611 -- 617 (2017).

\bibitem{Lustig2022}
E.~Lustig, L.~J. Maczewsky, J.~Beck, T.~Biesenthal, M.~Heinrich, Z.~Yang,
  Y.~Plotnik, A.~Szameit, and M.~Segev, \enquote{Three-dimensional photonic
  topological insulator induced by lattice dislocations,}
  {{arXiv:2204.13762 [physics.optics]}}  (2022).

\bibitem{Yang2020}
Z.~Yang, E.~Lustig, Y.~Lumer, and M.~Segev, \enquote{Photonic {F}loquet
  topological insulators in a fractal lattice,} {{Light
  Sci Appl}} \textbf{9} (2020).

\bibitem{Fu2020b}
Z.~Fu, N.~Fu, H.~Zhang, Z.~Wang, D.~Zhao, and S.~Ke, \enquote{Extended {SSH}
  model in non-{H}ermitian waveguides with alternating real and imaginary
  couplings,} {{Applied Sciences}} \textbf{10} (2020).

\bibitem{Zeuner2015}
J.~M. Zeuner, M.~C. Rechtsman, Y.~Plotnik, Y.~Lumer, S.~Nolte, M.~S. Rudner,
  M.~Segev, and A.~Szameit, \enquote{Observation of a topological transition in
  the bulk of a non-{H}ermitian system,} {{Phys. Rev.
  Lett.}} \textbf{115}, 040402 (2015).

\bibitem{Weidemann2020}
S.~Weidemann, M.~Kremer, T.~Helbig, T.~Hofmann, A.~Stegmaier, M.~Greiter,
  R.~Thomale, and A.~Szameit, \enquote{Topological funneling of light,}
  {{Science}} \textbf{368}, 311--314 (2020).

\bibitem{Meier2018}
E.~J. Meier, F.~A. An, A.~Dauphin, M.~Maffei, P.~Massignan, T.~L. Hughes, and
  B.~Gadway, \enquote{Observation of the topological {A}nderson insulator in
  disordered atomic wires,} {{Science}} \textbf{362},
  929--933 (2018).

\bibitem{Stutzer2018}
S.~Stützer, Y.~Plotnik, Y.~Lumer, P.~Titum, N.~H. Lindner, M.~Segev, M.~C.
  Rechtsman, and A.~Szameit, \enquote{Photonic topological {A}nderson
  insulators,} {{Nature}} \textbf{560}, 461 -- 465 (2018).

\bibitem{Kraus2012}
Y.~E. Kraus, Y.~Lahini, Z.~Ringel, M.~Verbin, and O.~Zilberberg,
  \enquote{Topological states and adiabatic pumping in quasicrystals,}
  {{Phys. Rev. Lett.}} \textbf{109}, 106402 (2012).

\bibitem{zilberberg2018photonic}
O.~Zilberberg, S.~Huang, J.~Guglielmon, M.~Wang, K.~P. Chen, Y.~E. Kraus, and
  M.~C. Rechtsman, \enquote{Photonic topological boundary pumping as a probe of
  4{D} quantum {H}all physics,} {{Nature}} \textbf{553},
  59--62 (2018).

\bibitem{Wimmer2017}
M.~Wimmer, H.~M. Price, I.~Carusotto, and U.~Peschel, \enquote{Experimental
  measurement of the {B}erry curvature from anomalous transport,}
  {{Nature Phys}} \textbf{13}, 545 -- 550 (2017).

\bibitem{Kitagawa2012}
T.~Kitagawa, M.~A. Broome, A.~Fedrizzi, M.~S. Rudner, E.~Berg, I.~Kassal,
  A.~Aspuru-Guzik, E.~Demler, and A.~G. White, \enquote{Observation of
  topologically protected bound states in photonic quantum walks,}
  {{Nat Commun}} \textbf{3} (2012).

\bibitem{Lai1989a}
Y.~Lai and H.~A. Haus, \enquote{Quantum theory of solitons in optical fibers.
  {I}. time-dependent {H}artree approximation,} {{Phys.
  Rev. A}} \textbf{40}, 844--853 (1989).

\bibitem{Lai1989b}
Y.~Lai and H.~A. Haus, \enquote{Quantum theory of solitons in optical fibers.
  {II}. exact solution,} {{Phys. Rev. A}} \textbf{40},
  854--866 (1989).

\bibitem{Larre2015b}
P.-E. Larr\'e and I.~Carusotto, \enquote{Propagation of a quantum fluid of
  light in a cavityless nonlinear optical medium: General theory and response
  to quantum quenches,} {{Phys. Rev. A}} \textbf{92},
  043802 (2015).

\bibitem{Kane2002}
C.~L. Kane, R.~Mukhopadhyay, and T.~C. Lubensky, \enquote{Fractional quantum
  {H}all effect in an array of quantum wires,} {{Phys.
  Rev. Lett.}} \textbf{88}, 036401 (2002).

\bibitem{Teo2014}
J.~C.~Y. Teo and C.~L. Kane, \enquote{From {L}uttinger liquid to non-{A}belian
  quantum {H}all states,} {{Phys. Rev. B}} \textbf{89},
  085101 (2014).

\bibitem{Budich2017}
J.~C. Budich, A.~Elben, M.~\L{}\k{a}cki, A.~Sterdyniak, M.~A. Baranov, and
  P.~Zoller, \enquote{Coupled atomic wires in a synthetic magnetic field,}
  {{Phys. Rev. A}} \textbf{95}, 043632 (2017).

\bibitem{yuan2018synthetic}
L.~Yuan, Q.~Lin, M.~Xiao, and S.~Fan, \enquote{Synthetic dimension in
  photonics,} {{Optica}} \textbf{5}, 1396--1405 (2018).

\bibitem{Ozawa2016}
T.~Ozawa, H.~M. Price, N.~Goldman, O.~Zilberberg, and I.~Carusotto,
  \enquote{Synthetic dimensions in integrated photonics: From optical isolation
  to four-dimensional quantum hall physics,} {{Phys. Rev.
  A}} \textbf{93}, 043827 (2016).

\bibitem{Yuan2016}
L.~Yuan, Y.~Shi, and S.~Fan, \enquote{Photonic gauge potential in a system with
  a synthetic frequency dimension,} {{Opt. Lett.}}
  \textbf{41}, 741--744 (2016).

\bibitem{dutt_creating_2022}
A.~Dutt, L.~Yuan, K.~Y. Yang, K.~Wang, S.~Buddhiraju, J.~Vučković, and
  S.~Fan, \enquote{Creating boundaries along a synthetic frequency dimension,}
  {{Nature Communications}} \textbf{13}, 3377 (2022).

\bibitem{wang_multidimensional_2020}
K.~Wang, B.~A. Bell, A.~S. Solntsev, D.~N. Neshev, B.~J. Eggleton, and A.~A.
  Sukhorukov, \enquote{Multidimensional synthetic chiral-tube lattices via
  nonlinear frequency conversion,} {{Light: Science \&
  Applications}} \textbf{9}, 132 (2020). 

\bibitem{hu2020realization}
Y.~Hu, C.~Reimer, A.~Shams-Ansari, M.~Zhang, and M.~Loncar,
  \enquote{Realization of high-dimensional frequency crystals in electro-optic
  microcombs,} {{Optica}} \textbf{7}, 1189--1194 (2020).

\bibitem{Piccioli2021}
F.~S. Piccioli, A.~Szameit, and I.~Carusotto, \enquote{Topologically protected
  frequency control of broadband signals in dynamically modulated waveguide
  arrays,} {{Physical Review A}} \textbf{105}, 053519
  (2022).

\bibitem{Ozawa2021}
T.~Ozawa, \enquote{Artificial magnetic field for synthetic quantum matter
  without dynamical modulation,} {{Phys. Rev. A}}
  \textbf{103}, 033318 (2021).

\bibitem{Nemirovsky2021}
L.~Nemirovsky, M.-I. Cohen, Y.~Lumer, E.~Lustig, and M.~Segev,
  \enquote{Synthetic-space photonic topological insulators utilizing
  dynamically invariant structure,} {{Phys. Rev. Lett.}}
  \textbf{127}, 093901 (2021).

\bibitem{Hofstadter1976}
D.~R. Hofstadter, \enquote{Energy levels and wave functions of {B}loch
  electrons in rational and irrational magnetic fields,}
  {{Phys. Rev. B}} \textbf{14}, 2239--2249 (1976).

\bibitem{Chalopin2020}
T.~Chalopin, T.~Satoor, A.~Evrard, V.~Makhalov, J.~Dalibard, R.~Lopes, and
  S.~Nascimbene, \enquote{Probing chiral edge dynamics and bulk topology of a
  synthetic hall system,} {{Nat. Phys.}} \textbf{16},
  1017--1021 (2020).

\bibitem{datasheet}
M.~Polyanskiy, \enquote{Optical constants of {F}used silica (fused quartz),}
  \url{https://refractiveindex.info/?shelf=glass&book=fused_silica&page=Malitson}.
  Accessed: 20/04/2023.

\bibitem{ozawa2019topological}
T.~Ozawa and H.~M. Price, \enquote{Topological quantum matter in synthetic
  dimensions,} {{Nature Reviews Physics}} \textbf{1},
  349--357 (2019).

\bibitem{supmat}
{{See Supplemental Material}} .

\bibitem{Mukherjee2018b}
S.~Mukherjee, H.~Chandrasekharan, P.~Öhberg, N.~Goldman, and R.~Thomson,
  \enquote{State-recycling and time-resolved imaging in topological photonic
  lattices,} {{Nat Commun}} \textbf{9} (2018).

\bibitem{pelucchi2022potential}
E.~Pelucchi, G.~Fagas, I.~Aharonovich, D.~Englund, E.~Figueroa, Q.~Gong,
  H.~Hannes, J.~Liu, C.-Y. Lu, N.~Matsuda \emph{et~al.}, \enquote{The potential
  and global outlook of integrated photonics for quantum technologies,}
  {{Nature Reviews Physics}} \textbf{4}, 194--208 (2022).

\bibitem{Ozawa2017}
T.~Ozawa and I.~Carusotto, \enquote{Synthetic dimensions with magnetic fields
  and local interactions in photonic lattices,} {{Phys.
  Rev. Lett.}} \textbf{118}, 013601 (2017).

\bibitem{DeBernardis2023}
D.~De~Bernardis, F.~Piccioli, P.~Rabl, and I.~Carusotto, \enquote{Chiral
  quantum optics in the bulk a photonic quantum {H}all system,}
  {{arXiv}}:2302.14863 (2023).

\bibitem{esposito2021perspective}
M.~Esposito, A.~Ranadive, L.~Planat, and N.~Roch, \enquote{Perspective on
  traveling wave microwave parametric amplifiers,}
  {{Applied Physics Letters}} \textbf{119}, 120501 (2021).

\bibitem{carusotto2013quantum}
I.~Carusotto and C.~Ciuti, \enquote{Quantum fluids of light,}
  {{Reviews of Modern Physics}} \textbf{85}, 299 (2013).

\bibitem{carusotto2020photonic}
I.~Carusotto, A.~A. Houck, A.~J. Koll{\'a}r, P.~Roushan, D.~I. Schuster, and
  J.~Simon, \enquote{Photonic materials in circuit quantum electrodynamics,}
  {{Nature Physics}} \textbf{16}, 268--279 (2020).

\end{thebibliography}

\begin{thebibliography}{1}
\newcommand{\enquote}[1]{``#1''}

\bibitem{Grant2004}
I.~Grant and W.~Phillips, \emph{Electromagnetism} (Wiley, Chicester, 2004), 2nd
  ed.

\end{thebibliography}
\end{document}